# Influence of Processing Parameters on the Mechanical Properties of HPT-Deformed Nickel/Carbon Nanotube Composites**

By Andreas Katzensteiner, Timo Müller, Karoline Kormout, Katherine Aristizabal, Sebastián Suarez, Reinhard Pippan and Andrea Bachmaier*

[*]     Dr. A. Bachmaier, A. Katzensteiner, T. Müller, Dr. K. Kormout, Prof. R. Pippan
*Erich Schmid Institute of Materials Science*
*Austrian Academy of Sciences*
*Jahnstr. 12, A-8700 Leoben, Austria*
*E-mail: Andrea.Bachmaier@oeaw.ac.at*
         K. Aristizabal, Dr. S. Suarez
*Functional Materials, Department of Materials Science*
*Saarland University*
*Campus D3.3, D-66123 Saarbrücken, Germany*

[**]     *A. Katzensteiner and A. Bachmaier gratefully acknowledge the financial support by the Austrian Science Fund (FWF): I2294-N36. S. Suarez and K. Aristizabal gratefully acknowledge the financial support from DFG (Grant: SU911/1-1). Additionally, K. Aristizabal would like to thank the German Academic Exchange Service (DAAD) for financial support.*

Nickel/carbon nanotube (Ni/CNT) composites with varying amounts of CNTs are deformed by high-pressure torsion (HPT) at different deformation temperatures to high strains, where no further refinement of the Ni matrix microstructure is observed. Mean Ni grain sizes increase with increasing HPT deformation temperature, while the size of the CNT agglomerates is significantly reduced. Additionally, the distribution of the agglomerates in the metal matrix becomes more homogenous. To investigate the mechanical performance of the HPT-deformed composites, uniaxial tensile and compression tests are conducted. Depending on the HPT deformation temperature and the resulting microstructure, either brittle or ductile fracture occurs, and the ultimate tensile strength varies between 900 and 2100 MPa. Increased HPT deformation temperatures induce a decrease in the anisotropy of the mechanical properties, mainly caused by a shrinking of the CNT agglomerates. It is shown, that tuning the HPT deformation temperature is the key for optimizing both the microstructure and the mechanical performance.

# 1. Introduction

Processing of metal matrix nanocomposites (MMCs) by severe plastic deformation (SPD) has gathered much interest in the materials science community because of the possibility of this method to not only create a nanograined or ultrafine grained microstructure, but also to disperse the second phase particles homogenously throughout the metal matrix.[1] These second phase particles are known to improve the mechanical properties of pure metals, due to their ability to pin dislocations and grain boundaries as well as to act as reinforcing phase. In nanocrystalline materials, they also inhibit grain growth, which improves the thermal stability of SPD processed metals.[2]

Carbon nanotubes (CNTs) consist of rolled up sheets of graphene, either single or multilayered, and possess outstanding material properties, such as high specific strength and high thermal conductivity.[3] These properties make them promising candidates as reinforcing phase for MMCs,[4] and the strength and ductility of such CNT/metal composites are the subject of several studies, using severe plastic deformation (SPD) processes, such as high pressure torsion (HPT) or equal channel angular pressing (ECAP),[5-7] and other processing techniques, for example electrodeposition.[8-12] Contrary to other carbon allotropes, such as nanodiamonds, CNTs are known to form large agglomerates when dispersed in MMCs.[13] The breaking-up and dispersion of these agglomerates, which has a significant influence on the mechanical performance of the MMCs, is therefore of outmost importance for improving the properties of such composites.

An often used method to investigate the mechanical properties of HPT deformed bulk materials is tensile testing.[14-16] Tensile tests are used as a direct measurement of the yield strength and the ultimate tensile strength (UTS), the uniform elongation, the stress and strain at fracture, and the Young´s modulus. Due to the size limitations of HPT processed samples, special care has to be exercised on the fabrication of appropriate tensile test specimens with these dimensions, since proportional standards for tensile test specimens are only given for larger samples, as used in industrial environments.[17,18]

The majority of studies on the topic of CNT-reinforced metals come to the conclusion, that the addition of CNTs as second phase particles has a beneficial impact on the strength and ductility. Yoon et al.[5] showed, that the addition of 5 or 10 vol% of CNTs to pure Cu and a subsequent HPT deformation of up to 10 revolutions increases the UTS from 190 MPa to 352 and 345 MPa, respectively. A similar UTS increase has been reported by Asgharzadeh and Kim[7] in Al reinforced with 3 vol% CNTs and HPT deformed up to 10 revolutions. Other consolidation methods have also shown to improve the mechanical properties of CNT reinforced metals. Yang et al.[8] reported a 1.7 times higher UTS and an elongation about 18 % higher of ball-milled and extrusion-sintered Al containing 2.5 wt% CNTs compared to pure Al. Spark plasma sintered Cu mixed with 5 vol% CNTs showed a comparable strength increase.[9]

The mechanical properties of Ni, as metal basis in different carbon-based reinforced composites, have been the subject of several studies as well. An increase of the UTS by a factor of almost two without a decrease in ductility, compared to pure Ni, through the addition of carbon impurities between 0.008 and 0.06 wt% and subsequent HPT deformation, has been reported by Rathmayr and Pippan.[16] On the other hand, a study by Archakov et al.[19] revealed a decrease in the ductility, but no increase in the UTS in ball milled and sintered Ni powder with 0.5 to 2 wt% of thermally expanded graphite, compared to pure Ni. Dai et al.[12] showed a hardness and UTS increase of CNT reinforced electrodeposited Ni, compared to the pure metal without significant loss of ductility.

A previous study of our group has shown, that the HPT deformation temperature has a strong influence on the hardness of Ni/CNT composites with up to 3 wt% CNT content and also on the dispersion of the CNTs in the Ni matrix.[20] Therefore, a two-temperature deformation process has been developed to optimize hardness, grain size, CNT agglomerate size and CNT distribution in the metal matrix of these composites.

In this study, the influence of the HPT deformation temperature on the strength and ductility of Nickel composites, with up to 3 wt% of CNTs has been investigated by tensile and compression tests, in order to optimize the mechanical performance. While tensile tests[21,22] and ultrasonic velocity measurements[23] have been used in previous studies to investigate the mechanical anisotropy of SiC reinforced MMCs, it is, to the best of our knowledge, the first time that compression tests have been used to investigate the anisotropy of HPT deformed Ni/CNT MMCs and the influence of deformation temperature and CNT distribution on it.

## 2. Experimental

Ni/CNT composites were produced via colloidal mixing of Ni dendritic powder (Alfa Aesar, -325 mesh) with Multiwall CNTs (Graphene Supermarket, USA, purity > 95 %, individual particle diameter of 20–85 nm, mean length between 5-15 µm) in an ethylene glycol dispersion according to ref.[13] with final CNT weight percentages of 0.1, 0.25, 0.5, 1, 2 and 3. These weight percentages correspond to volume percentages of 0.42, 1.04, 2.08, 4.17, 8.33 and 12.5. The composite samples had a diameter of 8 mm and a height of about 1 mm. HPT was used at a velocity of 0.6 revolutions per minute and a hydrostatic pressure of 7.5 GPa, to obtain a saturated microstructure, as described in ref.[24] The varied processing parameters were the deformation temperature and the number of revolutions, deforming either for 30 revolutions at room temperature, 200, 300 and 400 °C or using a two-temperature deformation process with 30 revolutions at 400 °C and subsequent 10 revolutions at 200 °C. From each HPT sample, 2 tensile test specimens were produced with a recently developed fabrication method for round, small-scale tensile specimens, which allowed a high-accuracy fabrication with negligible material change.[18,25] Tensile tests of samples HPT-deformed at room temperature were not conducted in this study, because of the high disposition of such samples to develop cracks during deformation, which excluded them from valid tensile testing. The gauge length was 2.5 mm and the diameter was between 300 and 500 µm. Hence, a ratio of gauge length to gauge diameter of 5:1 to 8:1 was achieved. The tensile axis was set to be about 2 mm from the center of the HPT disc to ensure full microstructural saturation in the testing region and long enough pulling shoulders on both sides to avoid a pullout of the specimens from the grips. Tensile tests were then performed on a Kammrath & Weiss tensile stage with a 2 kN load cell and a testing velocity of 2.5 μm/s at room temperature. The complete tensile test setup and the evaluation software are thoroughly described in ref.[18,25] After tensile testing, the fracture surfaces were investigated with a scanning electron microscope (SEM) type LEO 1525 equipped with an In-lens detector.

For the compression tests, 3 small cubes with side lengths of about 1 mm were cut from each HPT sample at a radius of about 3 mm from the disc center and polished with fine abrasive paper to get smooth, parallel surfaces. These cubes were then each compressed along one of the three directions, as depicted in **figure 1** (axial, radial and tangential), in the Kammrath & Weiss tensile/compression module with a 10 kN load cell and a testing velocity of 0.5 μm/s at room temperature. Test setup and data evaluation were analogous to the tensile tests. The microstructure of the 2 wt% CNT samples was investigated with SEM using a back scattered electron (BSE) detector and with a transmission electron microscope (TEM) JEOL 2100F. The SEM images were recorded in tangential direction at a radius of 3 mm and the TEM images were recorded in axial direction at a radius of 2 mm. To determine the anisotropy of the microstructure, additional micrographs of the HPT samples were recorded with SEM in axial and radial direction at a radius of 3 mm.

### 3. Results and Discussion

### 3.1. Microstructural development

The influence of the deformation temperature on the microstructure and the hardness of Ni/CNT MMCs has been described in ref.[20] It was shown, that the HPT deformation temperature has a strong influence on the hardness, the grain size of the Ni matrix and the agglomerate sizes of the CNTs and their distribution in the metal matrix. **Figure 2** exemplarily shows the microstructural change of a 2 wt% Ni/CNT composite with increasing HPT deformation temperature. HPT at 200 °C results in large, irregular CNT agglomerates, imbedded inhomogeneously in a nanograined Ni matrix (figure 2 a). Increasing the HPT deformation temperature to 300 °C and 400 °C (figures 2 b and c, respectively) leads to an increase in the Ni matrix grain size, a decrease in the size of the CNT agglomerates and a homogenization of the CNT distribution in the matrix. A two-temperature deformation process has been shown to optimize the microstructure, resulting in a Ni matrix with grain sizes smaller than 100 nm and mostly small, homogenously distributed CNT agglomerates (figure 2 d). Some larger agglomerates are still present in this sample due to the high amount of CNTs. **Figure 3 a** shows the corresponding hardness evolution of these samples, with a decrease in the saturation hardness with increasing HPT deformation temperature from 702 HV down to 458 HV and a hardness increase up to 855 HV of the sample deformed at two temperatures.

**3.2. Mechanical properties**

The results of the tensile tests reveal a strong influence of the deformation temperature on the mechanical properties of the composites. **Figure 3 b** exemplarily shows this for the specimens with 0.25 wt% and 2 wt% CNT content. Both specimens HPT-deformed at a temperature of 200 °C, show fracture in the elastic regime, with a fracture stress of 2121 MPa and 2066 MPa, respectively. At 300 °C, the 0.25 wt% specimen shows a stronger decrease in the UTS down to 1338 MPa, and a higher fracture strain of 4 %, than the 2 wt% specimen with 1558 MPa UTS and 0.5 % fracture strain. A further increase in the deformation temperature to 400 °C decreases the UTS to 892 MPa on the 0.25 wt% specimen, and to 1221 MPa on the 2 wt% specimen. The ductility increases to 4.9 % and 2.1 % fracture strain, respectively. The UTS and ductility of the specimen with low CNT content, deformed with the two-temperature process, is relatively high (1671 MPa with 3.1 % fracture strain). The specimen with high CNT content fails in the elastic regime, even after deformation with the two-temperature process.

**Figure 4** displays UTS, yield strength, uniform elongation and reduction in area of the tensile tests of all CNT weight percentages and deformation temperatures. Specimens with more than 1 wt% CNTs, deformed at 200 °C or with the two-temperature process, failed in the elastic regime and therefore showed no plastic yield. In this case, the fracture stress is plotted instead of the UTS. The UTS (figure 4 a) and the yield strength (figure 4 b) show no clear relationship to the amount of CNTs. A similar trend was observed from hardness measurements, where no significant increase in the hardness with increasing CNT content was determined.[20] Another influencing factor is the HPT deformation temperature. The 400 °C specimens show the lowest UTS and yield strength values. Higher values are obtained for the specimens deformed at 300° in accordance with hardness measurements.[20] For the other temperatures (200°C and the two-temperature process), no clear temperature influence on UTS and yield strength for specimens < 1 wt% CNTs can be discerned, but generally the highest values are achieved. The uniform elongation and reduction in area are plotted in figure 4 c and d, respectively. At low CNT contents, there is no evident influence of the HPT deformation temperature on the uniform elongation. The uniform elongation is, however, influenced by the deformation temperature at higher CNT contents, where it is considerably higher after deformation at 400°C compared to deformation at 300°C. The reduction in area (figure 4 d), as a second measure of ductility, is more dependent on the CNT content than on the HPT deformation temperature and decreases strongly with increasing CNT content. Furthermore, the reduction in area is generally lower for higher CNT contents at all HPT deformation temperatures.

To evaluate the fracture surface of the tensile specimens, SEM micrographs were obtained. **Figure 5** exemplarily shows the fracture surfaces of the 2 wt% specimens, HPT deformed at a) 200 °C, b) 300 °C, c) 400 °C and d) 400 + 200 °C. The specimen deformed at 200 °C shows a brittle fracture surface, while higher deformation temperatures resulted in the typical cup-and-cone micro-ductile fracture. The specimen deformed with the two-temperature process also shows a brittle fracture surface. High-magnification images, however, reveal the presence of dimples even in the specimens with generally brittle fracture surfaces. The size of those dimples seems to vary with the deformation temperature, and therefore with the grain size.

Depending on the CNT content, the fracture surfaces show a varying amount of large and irregular inclusions in specimens deformed at 200 °C and small, more disc-like inclusions in specimens deformed at higher temperatures (as indicated by arrows in figure 5). These features are most likely CNT agglomerates.

Compared to tensile test results of pure Ni[15] and Ni mixed with 0.5 – 2 wt% thermally expanded graphite (TEG), which was ball-milled to create a nickel-graphene-graphite composite,[19] the combination of Ni with high amounts of CNTs leads to an increase of the UTS, but also to a decrease in ductility even at higher HPT deformation temperatures. These results are also in contrast to the findings of ref.[16], where the addition of carbon impurities between 0.06 and 0.008 wt% to HPT deformed Ni were not enough to cause embrittlement due to carbon content at the grain boundaries. Similar results were obtained with Ni/CNT composites fabricated by electrodeposition.[12] The differences might be due to the higher carbon content in the present study, and the agglomerates, which could not be completely dissolved during HPT.

In ref.[16], the carbon content, as well as the HPT deformation temperature controlled the final saturation grain size. The hardness, which increased with decreasing grain size from 300 HV at a grain diameter of 400 nm, to 550 HV at a grain diameter of 120 nm, follow the Hall-Petch relation. Additionally, a linear trend was obtained if the hardness was plotted as function of the fourth root of the carbon content, which reflects the dependence of the grain boundary mobility on the carbon content.[16] In this study, specimens from samples deformed at room temperature could not be included in the tensile test results due the often observed crack formation during HPT processing or their high premature failure rate during tensile testing. Also samples deformed at 200°C and at two temperatures display a large scatter in the tensile test data due to their brittle failure in the elastic regime. To compare our results with ref.[16], the hardness for all CNT contents and HPT-deformation temperatures is plotted as function of grain size in **figure 6 a**. The mean grain sizes were measured from TEM images for the 2 wt% samples and from SEM images for all other samples, using the equivalent diameter method. Additionally, the data points available from ref.[16] are displayed. A linear relation with a similar slope as in ref.[16] with small deviations at smaller grain sizes is obtained for the Ni/CNT MMCs of this study. A Hall-Petch relationship between the grain size and the hardness was also found by Suarez et al.[29] in CNT-reinforced nickel at larger grain sizes, by Kim et al.[9] in CNT-reinforced copper, and by Choi et al.[11] in CNT-reinforced aluminum. It thus strengthens the argument, that the hardness increase of MMCs with CNTs as reinforcing phase is the result of a decreased grain size and the grain boundaries being obstacles for dislocation motion. Nonetheless, there can be other mechanisms at play, such as the dislocation density (work hardening) and the particle dispersion (Orowan strengthening). The latter strengthening mechanism is only valid for CNT agglomerates being smaller than the Ni matrix grain size and thus there is only a small or even negligible contribution to the strength of the composites in this study. CNT agglomerates larger than the grain size do not contribute to the strength substantially.

In **figure 6 b**, the hardness is plotted as a function of the carbon content for CNT samples deformed at room temperature and at 400°C and compared to the hardness data of ref.[16] (only data for room temperature and 400°C available). For both Ni/CNT MMCs, a linear relation is obtained as well. The slope of the curves are, however, considerably smaller. A decreased slope suggests a smaller influence of carbon on the saturation grain size after HPT deformation.[16] In ref.[16], the carbon is mainly enriched along the grain boundaries and has thus a large effect on the final grain size after HPT deformation. In ref.[30], the evolution of the CNT distribution homogeneity is thoroughly analyzed and different size descriptors and shape factors for the CNT agglomerates were obtained. It was shown that the diameter and the nearest neighbor distance of the CNT agglomerates decreased with increasing strain, but the CNT area fraction did not change significantly with increasing strain values. Thus, it is assumed that the carbon in this study is mostly present as CNTs in the agglomerates as well and might be only to a small part enriched at the grain boundaries. However, a ratio between the carbon present in CNT agglomerates and at the grain boundaries cannot be determined.

The CNT content has, however, a bigger influence on the ductility of the composites than on the hardness, the UTS and the yield strength, especially after deformation at two temperatures, where low (< 1 wt%) amounts of small CNT agglomerates lead to a combination of relatively high ductility and high UTS, whereas high (≥ 1 wt%) CNT content specimens show a generally brittle behavior. Since pure Ni shows ductile fracture even for nanoscaled grain sizes,[15] a possible explanation for these results is, that the ductility is mainly governed by the number of CNT agglomerates which serve as micro defects in the Ni matrix. Due to the high amount of CNTs not all agglomerates can be reduced in size. Their size, shape and concentration are thought to be one of the main criteria for the type of fracture (c.f. figures 2 and 5). Especially the large, irregular CNT agglomerates can act as fracture nuclei due to being a bundle of mechanically interlocked CNTs with high strength, which are then strongly compressed and damaged during HPT deformation.[13]

In the region near the UTS, these micro defects grow and build up pores. Between these larger pores, small pores in the Ni matrix develop as consequence of the multi-axial stress distribution. Both types of pores subsequently grow until final fracture occurs. The size of the two different kinds of pores of the samples deformed at 200 °C and 400 + 200 °C (figures 5 a and d) is clearly different, whereas the fracture surfaces of the samples deformed at 300 °C and 400 °C display pores with a more comparable size (figures 5 b and c). This could be a consequence of the different microstructure (cf. figure 2), where the samples deformed at 200 °C and 400 + 200 °C show CNT agglomerates larger than the Ni grain size, and the samples deformed at 300 °C and 400 °C show CNT agglomerates smaller or the same size as the Ni grains. Similar behavior has been found in Al/SiC MMCs, where the fracture surfaces of tensile specimens show a brittle fracture of the SiC particles in a dimpled fracture of the Al matrix.[22]

**3.3. Anisotropy**

Our results regarding the mechanical anisotropy obtained by compression testing are shown in **figure 7**. The compressive stress is plotted as a function of the compressive strain of specimens with 0.25 and 2 wt% CNTs, HPT deformed at a) 200 °C, b) 300 °C, c) 400 °C and d) 400 + 200 °C. After HPT deformation at 200 °C, both compositions show low ductility, high yield stresses of up to 3000 MPa and a distinct anisotropy, indicated by a strong disparity of the yield stress between the three tested orientations (figure 7 a). Increasing the HPT-deformation temperature to 300 °C increases the ductility and decreases the yield stress for both specimens. The anisotropy also decreases, showing a similar decline in both specimens (figure 7 b). Only in radial direction, the 2 wt% specimen shows a different behavior. At 400 °C HPT deformation temperature, both specimens show a further decrease of the yield stress down to about 1100 MPa and a high ductility similar to the 300 °C specimens. The 2 wt% specimen still shows some anisotropy (figure 7 c). The specimens obtained by the two-temperature deformation process, shown in figure 7 d, reveal a clear disparity of the mechanical behavior of low-CNT specimens and high-CNT specimens. While the 0.25 wt% specimens show a high yield stress and ductility in combination with a low anisotropy, the 2 wt% specimens display a very brittle behavior with a high maximum stress. The fracture mode during compression clearly depends on the CNT content. Specimens with low CNT content deform either ductile or by shear band formation and strain localization at an angle of about 45 ° depending on the HPT deformation temperature. High CNT content specimens on the other hand fail generally through brittle fracture.

To assess how much the size and shape of the CNT agglomerates play a role in the anisotropy of the composites, SEM images of the microstructure have been taken in tangential, radial and axial direction at a HPT disc radius of 3 mm. **Figure 8** shows these images for samples with 2 wt% CNTs. At the lowest HPT deformation temperature (figures 8 a, b, c), large, irregular CNT agglomerates are inhomogenously dispersed in a fine grained Ni matrix. These agglomerates are mainly stretched in shear direction (figures 8 b, c), but also slightly elongated in radial direction (figure 8 a). An increase of the deformation temperature to 300 °C leads to an increase in the mean grain size of the Ni matrix (figures 8 d, e, f) and to a decrease of the mean agglomerate size, which are also more homogeneously distributed while still elongated in shear direction. Figures 8 g, h and i show an increase in the Ni matrix grain size at an HPT deformation temperature of 400 °C, while the agglomerate size and elongation is comparable to the samples deformed at 300°C. The SEM images of the sample deformed at two temperatures (400 and 200 °C) show a Ni matrix with very small grains and small, evenly distributed CNT agglomerates (figures 8 j, k, l). These CNT agglomerates are, however, elongated in the shear direction as well.

A decrease of the anisotropy of the mechanical properties, especially of the ductility, with increasing deformation temperature is observed in the compression tests, although the CNT agglomerates are elongated in shear direction in all investigated samples (figure 7). This can be attributed to the mean size decrease of elongated CNT agglomerates. A comparison between figures 7 and 8 shows accordance between the decrease of the anisotropy and the decrease of the agglomerate sizes. Because of the high CNT content in the 2 wt% sample, some larger agglomerates, randomly distributed in the matrix, remain even after deformation at higher temperatures (300°C and 400°C). This explains the higher anisotropy of the 2 wt% specimens compared to the 0.25% wt% specimen at these deformation temperatures (figure 7). Similar results were obtained by Ganesh and Chawla[22] with SiC reinforced Al MMCs. They showed, that an increase in the volume fraction of the reinforcing particles increases the anisotropy of the MMCs during tensile tests.

HPT deformed microstructures often consist of elongated grains, which are aligned nearly parallel to the shear plane and display a shear texture. An orientation dependency of the mechanical properties in Ni/CNT MMCs is also caused by the grain shape (grain aspect ratio) and the shear texture of the grains of the Ni matrix. In a previous study, TEM images of a Ni/CNT MMC sample with 1 wt% CNTs, HPT deformed for 30 revolutions at 400 °C and additional 5 revolutions at room temperature, show an elongation of the Ni grains in the Ni/CNT MMCs as well.[20] The influence of the grain aspect ratio and orientation on the anisotropy of HPT deformed Ni was already investigated by Rathmayr et al.[14] and cannot be entirely ruled out for the Ni/CNT MMCs. In ref.[14], however, no influence of the grain aspect ratio on the UTS and the uniform elongation was found and the observed difference in the UTS of about 100 MPa was attributed to the shear texture. If the shear texture plays a role in the compression strength of the investigated Ni/CNT MMCs, specimens oriented parallel to the shear plane (i.e. tangential testing direction) should have a higher compression strength than specimens oriented perpendicular to the shear plane (i.e. axial testing direction), [14] which is not observed (figure 7).

## 4. Summary

Ni/CNT MMCs with varying amounts of CNTs were HPT deformed at different temperatures and their mechanical performance was evaluated. Tensile tests showed that decreasing the deformation temperature has a stronger influence on the UTS of the composites than the CNT content. The UTS of specimens deformed with a two-temperature process was almost as high as that of the 200 °C specimens but provides a better ductility at low CNT contents. In general, the CNT content has a strong influence on the ductility, especially after the two-temperature deformation, where only specimens with 1 wt% and higher showed a brittle fracture behavior. An overall brittle fracture behavior was observed for specimens deformed at 200 °C, whereas ductile failure for specimens deformed at higher temperatures was observed. The difference in the fracture mode can be ascribed to the dependence of the size and shape of the CNT agglomerates on the deformation temperature. Compression tests showed that the anisotropy of mechanical properties decreases with increasing deformation temperature. Microstructural observations of all tested orientations suggest that the anisotropy measured in the compression tests is mainly determined by the size and shape of the CNT agglomerates.

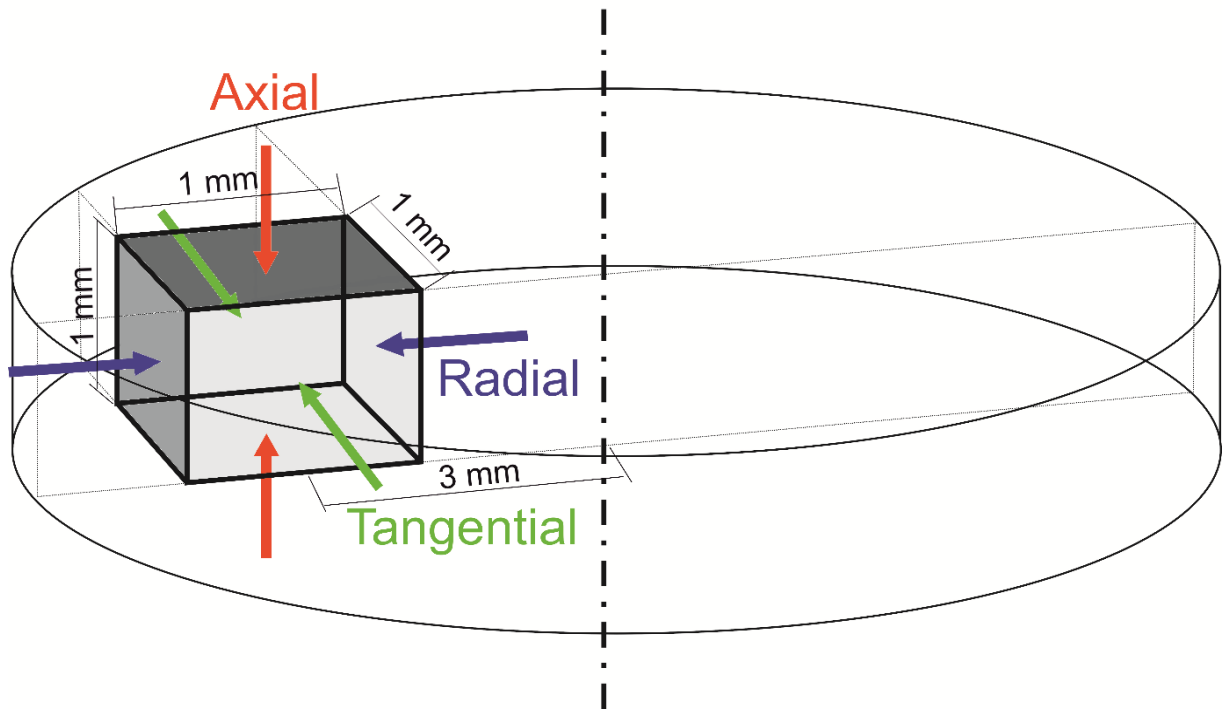

*Fig. 1. Schematic of a compression test specimen cut from an HPT disc with the orientations of the acting forces in each test direction.*

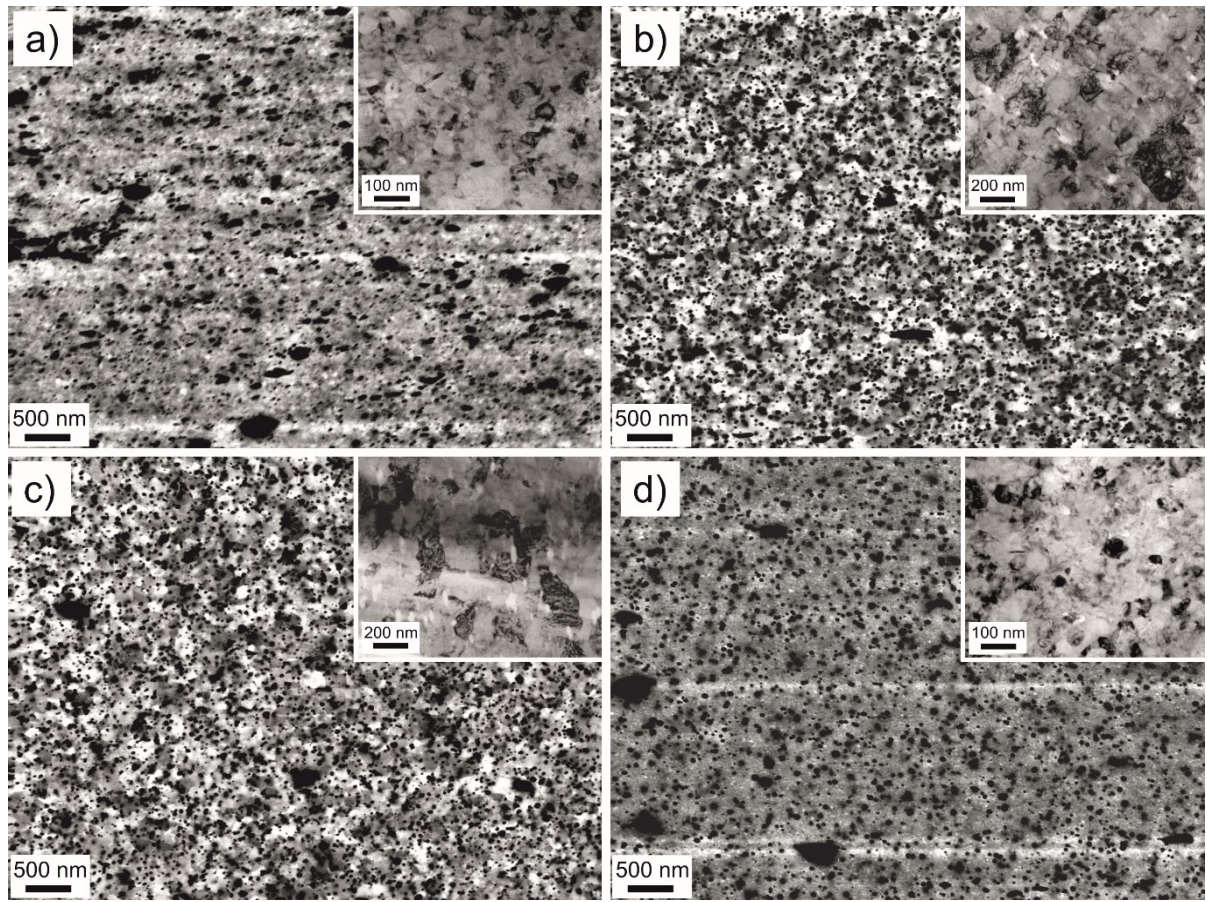

*Fig. 2. Microstructure images of 2 wt% Ni/CNT samples at a radius of 3 mm after HPT deformation at a) 200 °C, b) 300 °C, c) 400 °C and d) 400 + 200 °C (The inlays show TEM images taken with higher magnification at approximately the same position).*

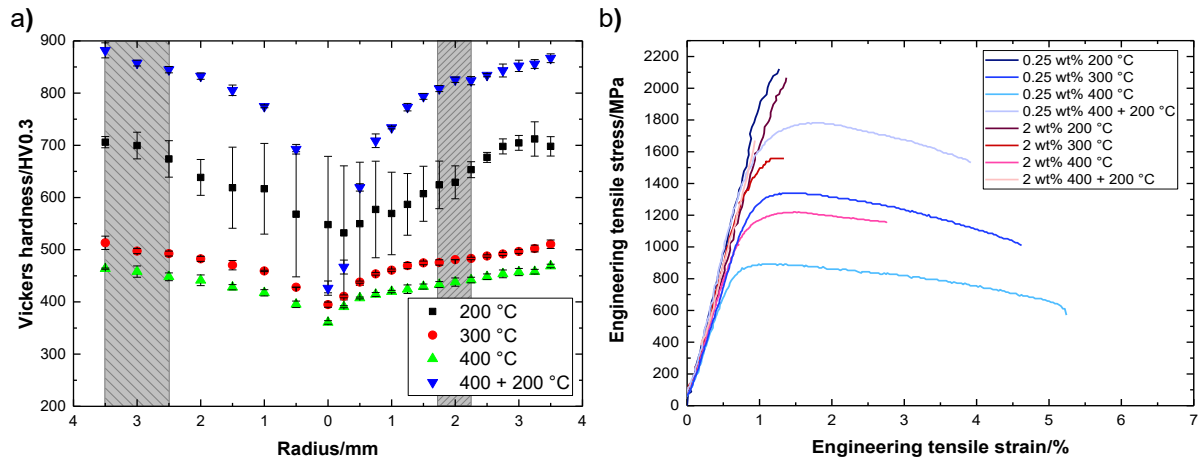

*Fig. 3. a) Vickers hardness as function of the HPT disc radius of Ni MMCs with 2 wt% CNTs, HPT deformed at 200 °C, 300 °C, 400 °C and 400 + 200 °C. The left and right gray areas indicate the compression region and the tensile test region, respectively. b) Engineering tensile stress-strain curves of Ni MMCs with 0.25 and 2 wt% CNTs, HPT deformed at 200 °C, 300 °C, 400 °C and 400 + 200 °C.*

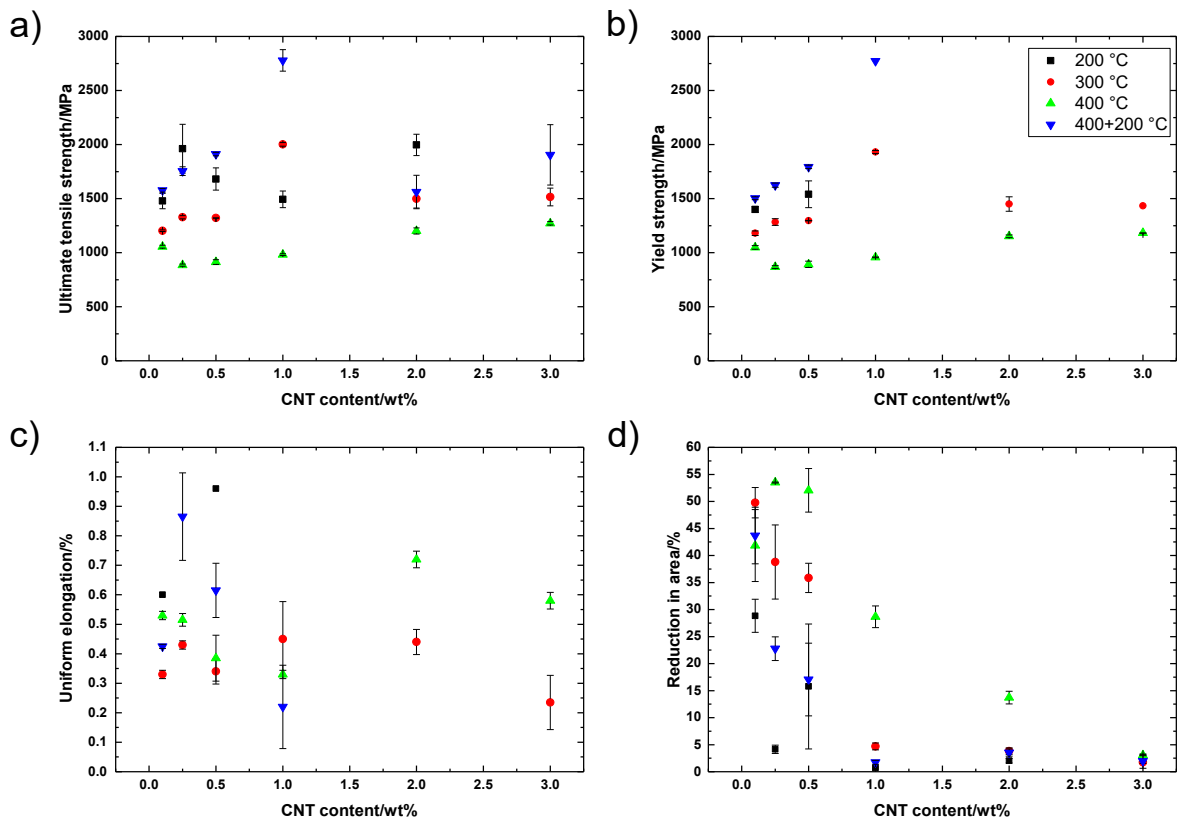

*Fig. 4. UTS (a), yield strength (b), uniform elongation (c) and reduction in area (d) as functions of the CNT content for specimens HPT deformed at different temperatures.*

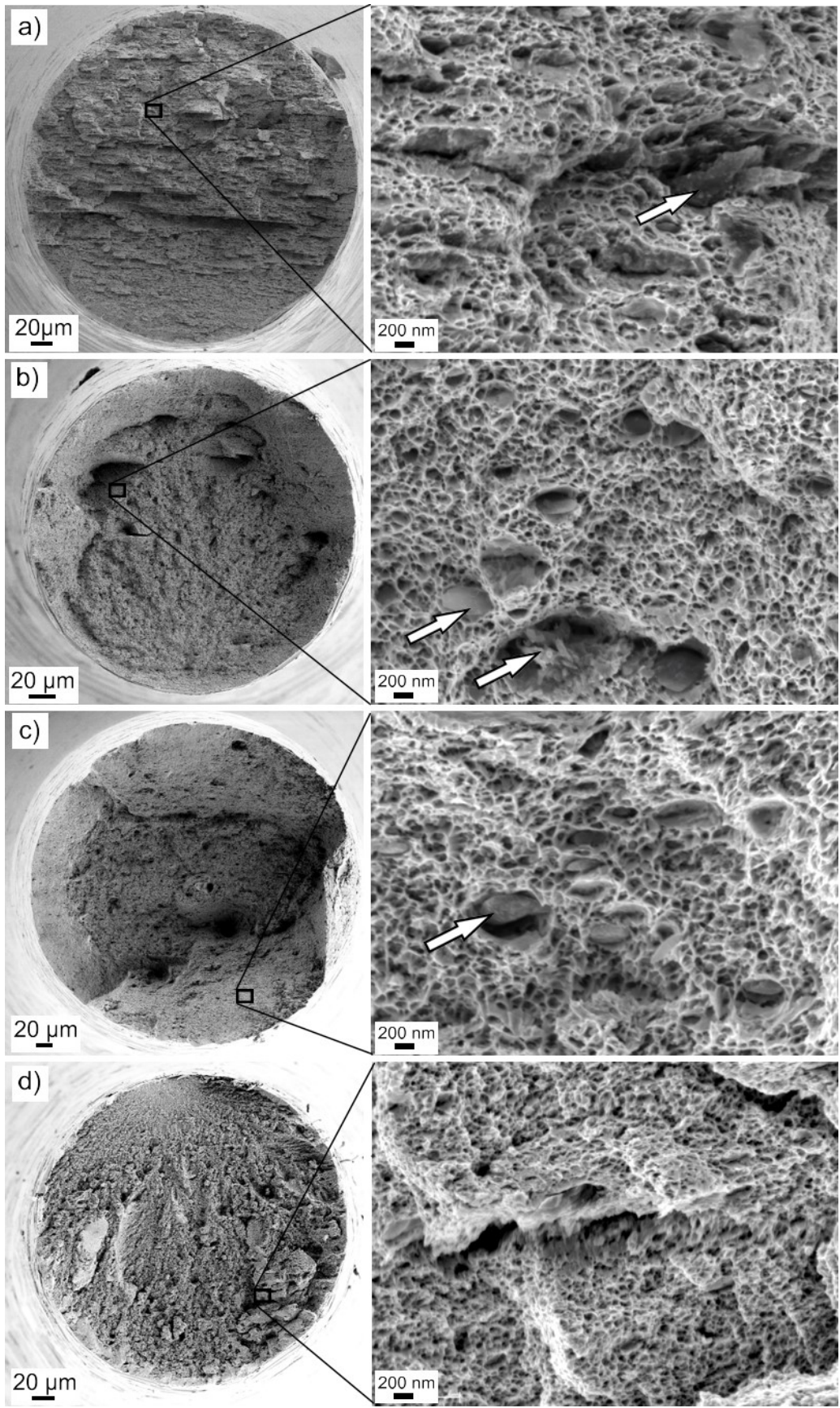

*Fig. 5. Fracture surfaces and detail fracture surfaces of Ni MMCs with 2 wt% CNTs HPT deformed at a) 200 °C, b) 300 °C, c) 400 °C and d) 400 + 200 °C. CNT agglomerates are marked by arrows.*

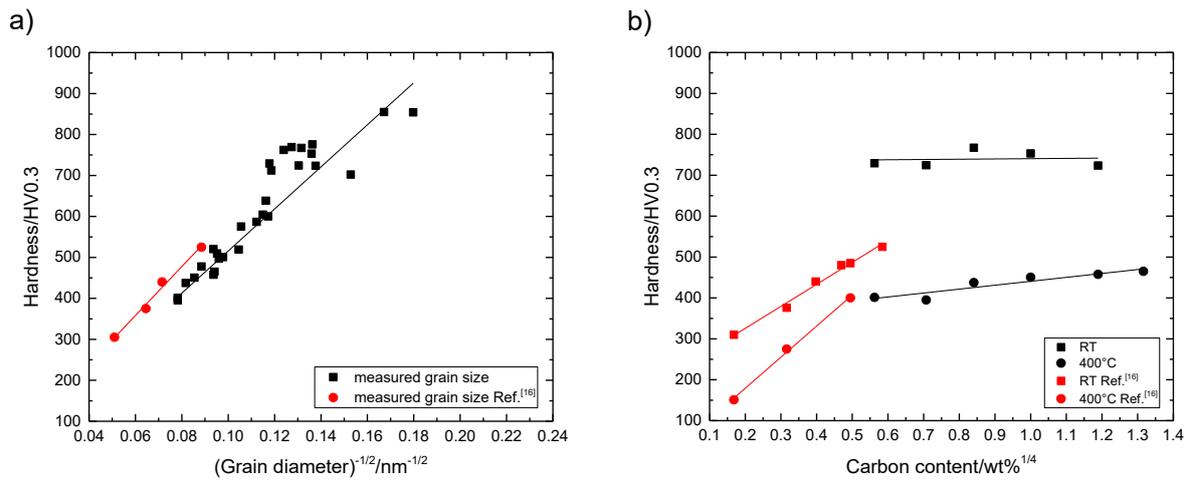

*Fig. 6. a) Saturation hardness as a function of grain size. The red data points are taken from ref.[16]. b) Saturation hardness as function of carbon content for two different HPT-deformation temperatures. The red data points are taken from ref.[16].*

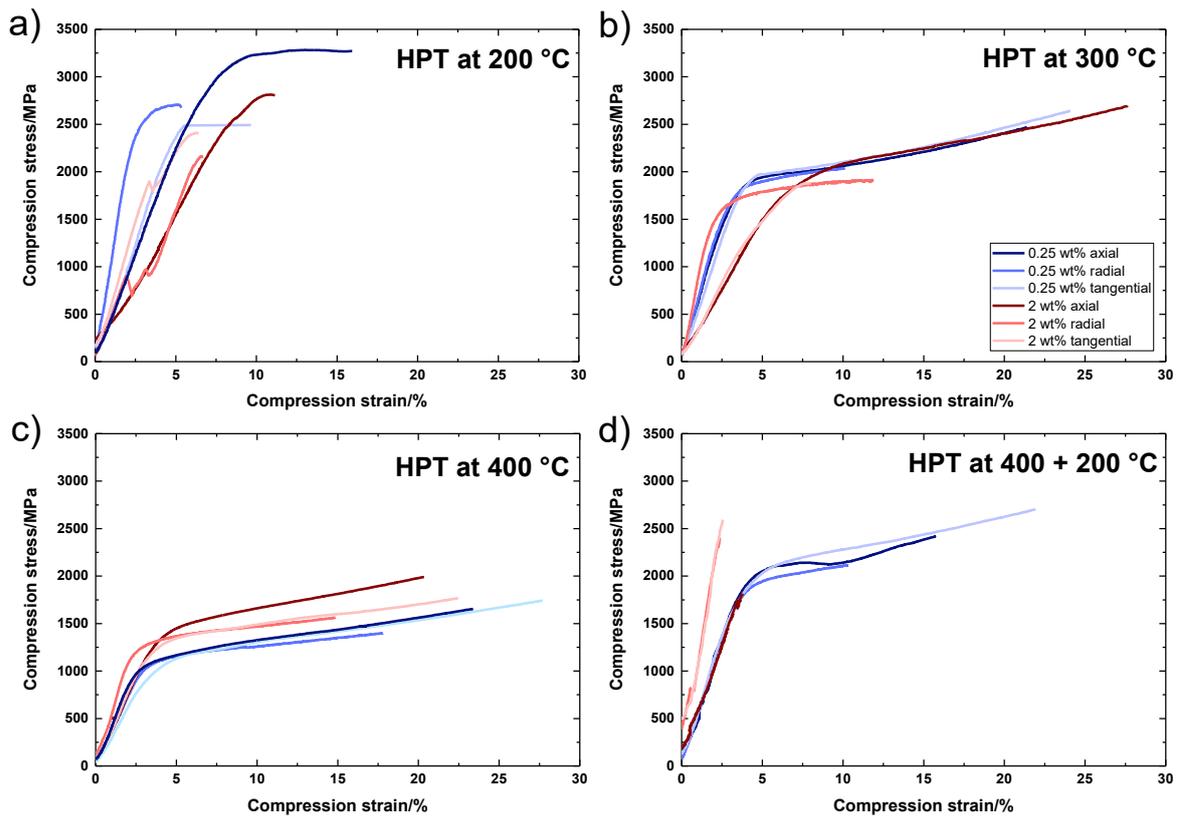

*Fig. 7. Compression stress-strain curves in axial, radial and tangential direction of Ni MMCs with 0.25 and 2 wt% CNTs, HPT deformed at a) 200 °C, b) 300 °C, c) 400 °C and d) 400 + 200 °C.*

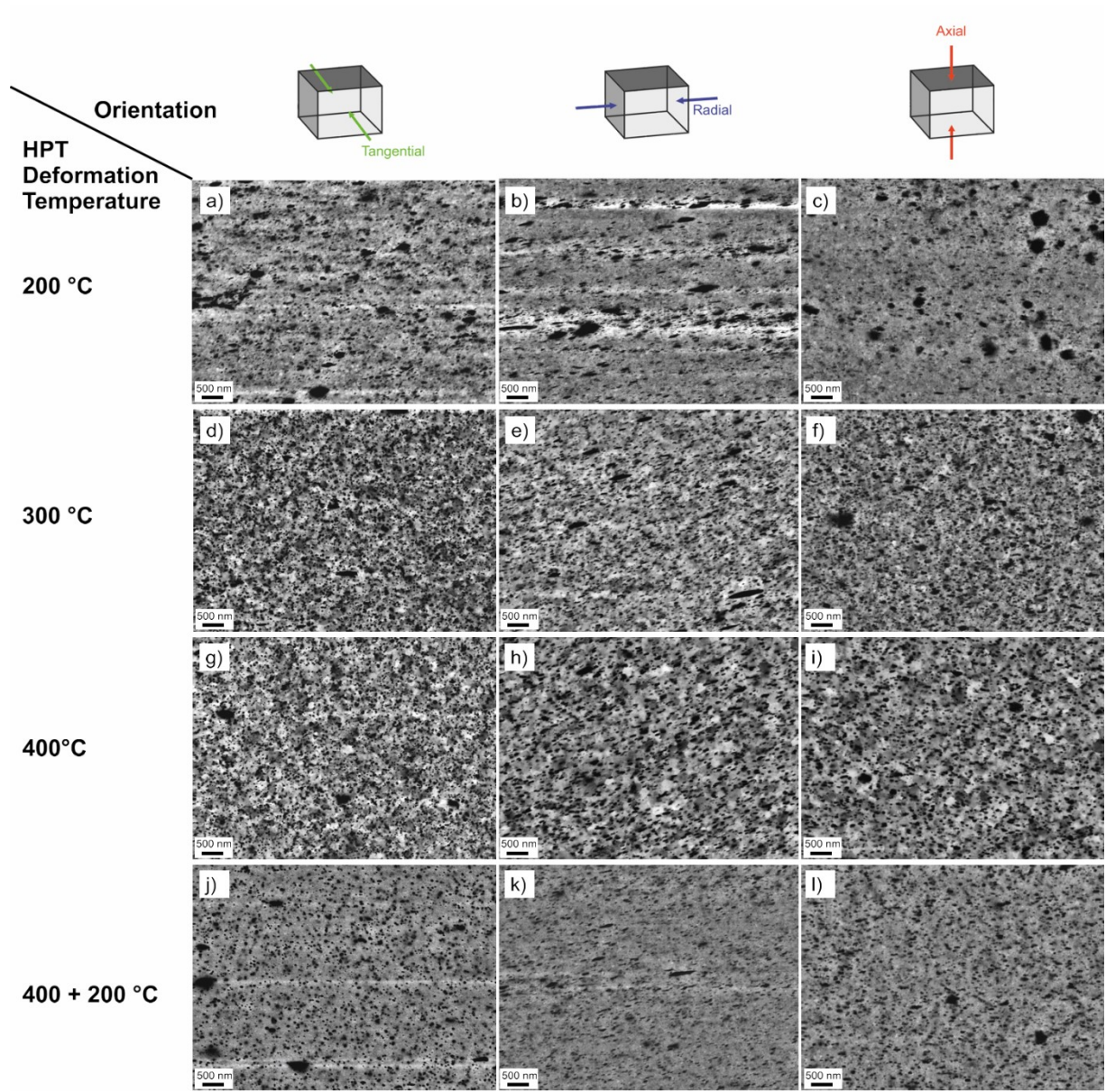

*Fig. 8. SEM images at 3 mm radius in tangential, radial and axial direction of Ni MMCs with 2 wt% CNTs, HPT deformed at 200 °C (a,b,c), 300 °C (d,e,f), 400 °C (g,h,i) and 400 + 200 °C (j,k,l).*

High pressure torsion (HPT) of Nickel with carbon nanotube (CNT) additives can lead to homogenous, nanograined metal matrix composites. The mechanical performance of such composites is investigated with tensile and compression tests for different CNT contents and HPT deformation temperatures. The HPT deformation temperature is the key to fabricate high strength Nickel/CNT metal matrix composites with reasonable ductility.

Andreas Katzensteiner, Timo Müller, Karoline Kormout, Katherine Aristizabal, Sebastián Suarez, Reinhard Pippan and Andrea Bachmaier*■<B>…<B>■

Influence of Processing Parameters on the Saturation Microstructure and Mechanical Properties of HPT-deformed Nickel/Carbon Nanotube Composites

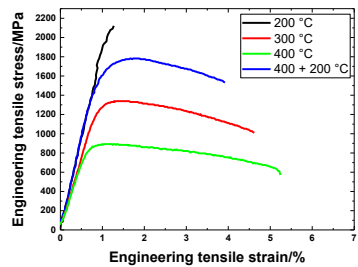